\documentclass[
reprint,
showpacs,
amsmath,amssymb,prd]{revtex4-1}

\usepackage{graphicx}
\usepackage{pstricks}
\usepackage{dcolumn}
\usepackage{bm}
\usepackage{esdiff}
\usepackage{hyperref}
\usepackage[mathlines]{lineno}

\newcommand{\powerset}{\raisebox{.15\baselineskip}{\Large\ensuremath{\wp}}}
\begin{document}

\title{Redshift and frequency comparison in Schwarzschild spacetime}

\author{Dennis Philipp, Eva Hackmann, Claus L\"ammerzahl}

\affiliation{ZARM, University of Bremen, 28359 Bremen, Germany}

\begin{abstract}
We derive exact expressions for the relativistic redshift between an Earth--bound observer, that is meant to model a standard clock on the Earth's surface, and various (geodesic) observers in the Schwarzschild spacetime. We assume that the observers exchange radial light signals to compare the frequencies of standard clocks, which they transport along their respective worldlines.

We calculate the redshift between an Earth--bound clock and static observers, observers in radial free fall, on circular geodesics, and on arbitrary bound quasi--elliptical orbits. For the latter case, we consider as examples an almost circular orbit, the Schwarzschild analog of Galileo satellites 5 orbits with a moderate eccentricity, and a highly elliptical orbit as special examples. Furthermore, we also use orbits close to a Schwarzschild black hole to highlight the influence of the relativistic perigee precession on the redshift signal. Calculating a post--Newtonian expansion of our results, the total redshift is decomposed into its special relativistic Doppler parts and the gravitational part due to the theory of General Relativity.

To investigate the impact of higher order relativistic multipole moments on the gravitational redshift, we consider static observers in a general Weyl spacetime. We give a general expression for the mutual redshift of their standard clocks and consider in particular the effect of the relativistic quadrupole as a modification of the Schwarzschild result.
\end{abstract}

\pacs{}
\keywords{}
\maketitle



\section{\label{Sec_intro} Introduction}

The theory of General Relativity (GR) predicts that the frequencies of clocks are influenced by the clocks' motion as well as their respective positions in the gravitational field. From Special Relativity (SR), the parallel and transverse Doppler effects on the redshift between moving observers are well--known. However, GR predicts yet another redshift effect related to gravity, i.e.\ to the spacetime curvature. 
Since only relative measurements are meaningful clocks shall be linked and their mutual redshift is to be determined, either by real optical fiber links when the clocks are close to each other or by exchanging electromagnetic signals. 

The first experiment in this respect was conducted by Pound and Rebka in 1960 \cite{Pound1960}. The authors verified the change of a photon's frequency during propagation in the gravitational field of the Earth. In 1966, the GEOS-1 satellite was used to observe the relativistic Doppler effects \cite{Jenkins1969}. To date, the best test of the gravitational redshift is still given by the Gravity Probe A (GPA) experiment, see \cite{Vessot1979,Vessot1980,Vessot1989}. The measurement was conducted in 1976 and used two hydrogen masers, one of which was carried to a height of $10^3\,$km by a Scout D rocket to be compared with the hydrogen maser on ground. To quote from the results of this seminal experiment \cite{Vessot1980}: 
\emph{
	``The agreement of the observed relativistic frequency shift with prediction is at the $70 \times 10^{-6}$ level.''
}

The gravitational redshift, however, was confirmed with an accuracy of $1.4 \times 10^{-4}$, see Ref.\ \cite{Vessot1989}.
At present, the ESA--funded study GREAT and DLR--funded study RELAGAL aim to improve the accuracy of the gravitational redshift test by analyzing clock data from the Galileo satellites 5 and 6, which were fortunately lunched into elliptic orbits \cite{Delva2015}. The authors claim that the data analysis and integration over one year can improve the GPA limit to about $4 \times 10^{-5}$. The upcoming Atomic Clock Ensemble in Space (ACES) will test the gravitational redshift at the $10^{-6}$ level \cite{ACES2011,Cacciapuoti2009}. 
In the past, several satellite test of SR and GR that involve space--based clocks have been proposed, see, e.g., \cite{Laemmerzahl2001, Dittus2007}. A recent approach to test the GR prediction of the gravitational redshift uses the RadioAstron satellite and the authors are confident to reach an accuracy in the $10^{-5}$ regime \cite{Litvinov2017}. Further prospects for future satellite missions using spacecraft clocks are discussed in Ref.\ \cite{Angelli2014}.

For a general overview of ''The Confrontation between General Relativity and Experiment``, we refer the reader to the living review article \cite{Will2006}. Details on relativistic effects in the Global Positioning System (GPS), in particular on timing and redshifts, can be found in the review in Ref.\ \cite{Ashby2003} and references therein. For time transfer in the vicinity of the Earth and definitions of different time scales we refer to Ref.\ \cite{Nelson2011}.

The relativistic redshift effects can also serve to test different theories of relativistic gravity and, therefore, to test GR \cite{Bondarescu2015, Jetzer2017}. For instance, scalar--tensor theories and their parametrized post--Newtonian form where considered in Ref.\ \cite{Schaerer2014}, in which the authors derived the difference to the GR redshift signals for artificial satellite orbits around the Earth.

All the articles mentioned so far do only take into account the first order post--Newtonian approximation of the full relativistic redshift, which contains the Doppler parts known from SR and the gravitational redshift, which is a genuine effect of GR. This consideration is certainly sufficient to meet present technological capabilities and to provide the framework for clocks in space with contemporary accuracies for frequency comparison. However, in this work we derive an exact expression for the general relativistic redshift in static simple spacetimes, in particular the Schwarzschild spacetime. To construct a framework for future high--precision experiments in a top--down approach we find it is necessary to investigate all notions in full GR first to enable a thorough understanding and an undoubtedly correct interpretation of measurement results.


\section{\label{Sec_setting} Setting}
\subsection{Geometry, notation and redshift definition}
\label{SubSec_Notation}
We use Einstein's summation convention, greek indices are spacetime indices and run from 0 to 3, and latin indices are purely spatial indices running from 1 to 3. Our metric signature convention is $(-,+,+,+)$. 

The Schwarzschild spacetime is the simplest solution of Einstein's vacuum field equation and describes the spacetime outside a spherically symmetric mass distribution. For the purpose of this work, it shall serve as an approximation of the spacetime outside the Earth.
The geometry of the Schwarzschild solution is described by the metric
\begin{align}
	\label{Eq_Notation:SchwarzschildMetric}
  	g = - f(r) \mathrm{d}t^2 + f(r)^{-1} \mathrm{d}r^2 + r^2 \left( \mathrm{d}\vartheta^2 + \sin^2 \vartheta \, \mathrm{d}\varphi^2 \right) \, ,
\end{align}
where we use spherical coordinates $(t,r,\vartheta,\varphi)$, and the metric function $f(r)$ is given by
\begin{align}
	\label{Eq_Notation:SchwarzschildFkt}
  	f(r) = 1-\dfrac{2m}{r} \, .
\end{align}
The Schwarzschild radius is $r_s :=2m$, and it is related to the SI-mass $M$ of the source by
\begin{align}
	m = \dfrac{GM}{c^2} \, ,
\end{align}
where $G$ is Newton's gravitational constant, and $c$ is the vacuum speed of light. For the Earth, the Schwarzschild radius is about $2m_\oplus \approx 8.8\,$mm. In the following, We use natural units such that $G=1$ and $c=1$, if not stated otherwise. However, to calculate the post--Newtonian limit of our results, we introduce SI units again.

In GR, there is a universal formula for the frequency $\nu$ of a light ray that a given observers measures. Let a light signal be described by a curve $\lambda(s)$, where $s$ is an affine parameter along its trajectory. Furthermore, let the tangent vector to the light ray's worldline be $k := \mathrm{d}\lambda/\mathrm{d}s$. For any such lightlike signal, we have $g(k,k) = 0$. Let an observer's worldline be described by $x^\mu(\tau)$, and its four-velocity ${\mathrm{d}x/\mathrm{d}\tau =: \dot{x} =: u}$. The proper time $\tau$ is defined by the requirement that $g(u,u) = -1$. Hence, the observer's worldline is parametrized by proper time and, therefore, its clock is a standard clock. Such standard clocks can indeed be characterized operationally as shown in Ref.\ \cite{Perlick1987}.

\begin{figure}
	\centering
	\includegraphics[width=0.49\textwidth]{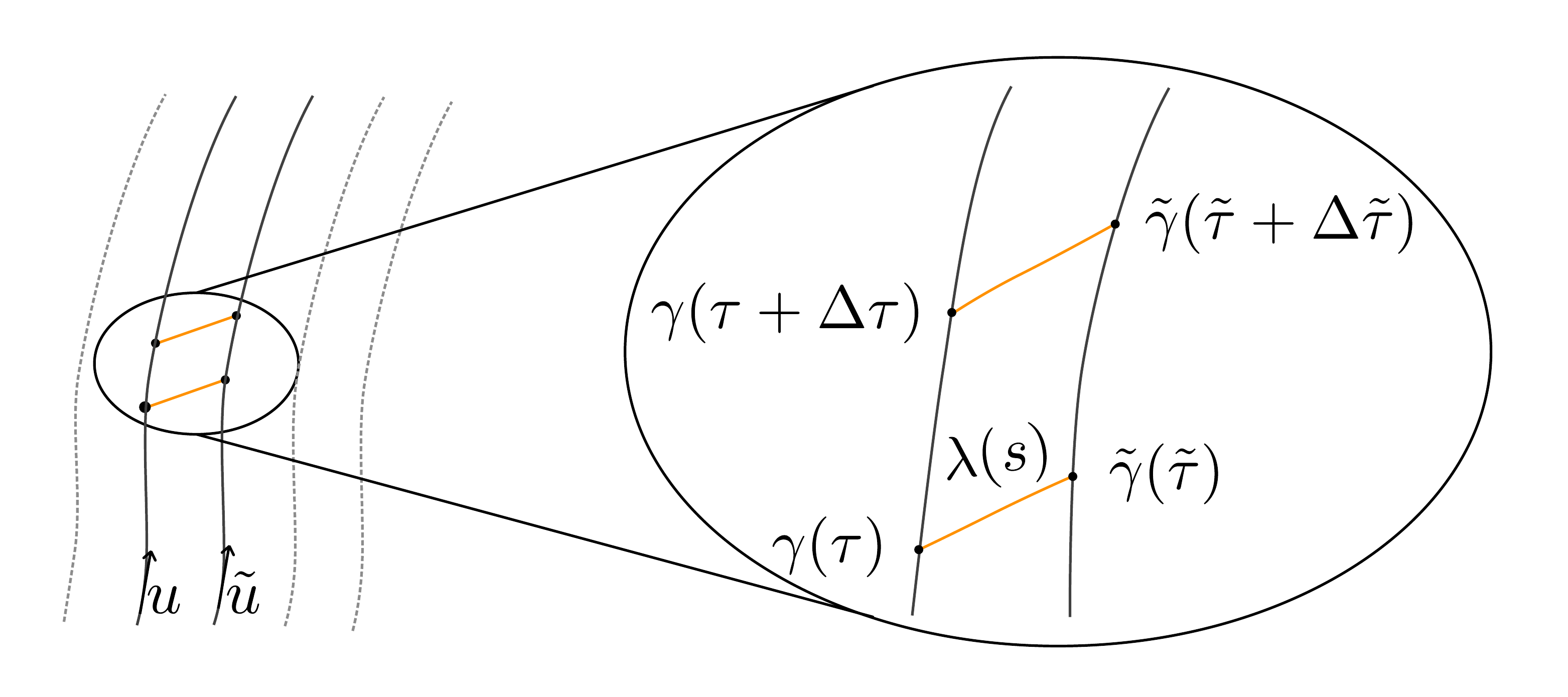}
	\caption{\label{Fig:redshiftSketch} Sketch of the signal transmission between emitter and receiver for the definition of the redshift. For any family of observers, we pick two wordlines $\gamma$ and $\tilde{\gamma}$ with tangent vectors $u$ and $\tilde{u}$, respectively. The continuous exchange of lightlike signals $\lambda$ gives rise to the definition of the redshift $z$ according to Eq.\ \eqref{Eq_Notation:RedshiftDef1}.}
\end{figure}

The frequency associated with the lightlike signal $\lambda$, which a given observer measures, is the scalar product (w.r.t.\ the metric) of the observer's four-velocity and the light ray's tangent vector, evaluated at the observers position $x$, i.e.\
\begin{align}
	\label{Eq_Notation:FrequDef}
	\nu =  g \big( k(x), u(x) \big) \, ,
\end{align} 
Now, we consider two different observers, which shall serve as emitter and receiver of a light signal sent from one to the other. Let the two observers be described by their respective four--velocities $u$ and $\tilde{u}$, with integral curves $\gamma(\tau)$ and $\tilde{\gamma}(\tilde{\tau})$ that are the observers' worldlines, respectively. 
We define the redshift $z$ between the two observers by
\begin{align}
	\label{Eq_Notation:RedshiftDef1}
	z+1 := \dfrac{\nu}{\tilde{\nu}} = \dfrac{d \tilde{\tau}}{d \tau} = \underset{{\Delta \tau \to 0}}{\mathrm{lim}}  \dfrac{\Delta \tilde{\tau}}{\Delta \tau} \, .
\end{align}
In GR, there is a universal formula for the redshift \cite{Kermack:1934}
\begin{align}
	\label{Eq_Notation:RedshiftDef2}
	1+z = \dfrac{\nu}{\tilde{\nu}} = \dfrac{g \big( k, u \big) \big|_{\gamma}}{g \big( k, \tilde{u} \big) \big|_{\tilde{\gamma}}} \, .
\end{align}
For the definition of the redshift according to Eq.\ \eqref{Eq_Notation:RedshiftDef1} see the sketch of the general situation in Fig.\ \ref{Fig:redshiftSketch}.

Equation \eqref{Eq_Notation:RedshiftDef2} allows to determine the redshift $z$ between two observers as a function of their respective worldlines and the exchanged light signal. The precise way of how this signal is exchanged from emitter to receiver still needs to be prescribed, i.e.\ the lightlike tangent vector $k$ needs to be specified. We will treat this issue in the next section.

In studies such as RELAGAL and GREAT \cite{Delva2015} however, usually clock residual are investigated. Hence, we also have to relate the proper times $\tau$ and $\tilde{\tau}$. Using Eqs.\ \eqref{Eq_Notation:RedshiftDef1} and \eqref{Eq_Notation:RedshiftDef2}, we obtain
\begin{align}
	\mathrm{d} \tilde{\tau} = \mathrm{d}\tau \, (z+1) \, ,
\end{align} 
and therefore
\begin{align}
	\tilde{\tau} = \int (z+1) \, \mathrm{d}\tau \, .
\end{align}
Thus, an analytic model of the redshift can be used to model the relation between the two proper times, and for a constant redshift we obtain the simple relation
\begin{align}
	\tilde{\tau} = (z+1) \, \tau \, .
\end{align}

\subsection{Observers and signal transfer}
We introduced two different kinds of observers, of which the worldlines $\gamma(\tau)$ and $\tilde{\gamma}(\tilde{\tau})$ are described as integral curves of their respective four-velocities $u$ and $\tilde{u}$. The worldline $\gamma$ shall now corresponds to a moving  (geodesic) observer which is the emitter of a signal, and $\tilde{\gamma}$ corresponds to an observer that we think of being the receiver attached to the Earth's surface. Within the approximation given by the Schwarzschild spacetime, the Earth is described by a sphere with radius $r_\oplus$. We assume $\tau$ and $\tilde{\tau}$ to be the observers' respective proper times. 
We can, without loss of generality, choose the equatorial plane to be the plane of motion. 
Even though we model the spacetime by the Schwarzschild geometry, we can of course consider observers fixed to the surface of a rotating Earth. This allows to accurately describe all Doppler contributions to the redshift but ignores all gravitomagnetic effects such as the clock effect, see \cite{Hackmann2014} and references therein.
Hence, we have in general
\begin{subequations}
	\label{Eq_Oberservs:4velocity}
	\begin{align}
		\label{Eq_}
  		(u^\mu) &= (u^t, u^r, u^\varphi, 0) \, , \\
  		(\tilde{u}^\mu) &= (\tilde{u}^t, 0, \tilde{u}^\varphi ,0) \, .
	\end{align}
\end{subequations}
Observers on the Earth's surface are fixed at spatial coordinates ${(r=r_\oplus,\,\vartheta=\pi/2)}$, and their worldlines are integral curves of the Killing vector field $\partial_t + \Omega \, \partial_\varphi$, where $\Omega$ is the Earth's angular velocity in natural units. Hence, we require that $\tilde{u}^\varphi = \Omega$ to have the Earth--bound observers forming a Born-rigid congruence. The value of $\Omega$ is roughly calculated by $\Omega \, c \approx 2\pi/86400\,$s. 

Both observers, $\gamma(\tau)$ and $\tilde{\gamma}(\tilde{\tau})$, shall exchange light signals to compare the frequencies of their standard clocks, which they transport along their respective worldlines. We assume that the frequency comparison is realized via radial lightlike geodesics $\lambda(s)$. Such a geodesic connects events on the wordlines of $\gamma$ and $\tilde{\gamma}$. For radial frequency comparison, the tangent vector $k$ of the light ray's worldline must lie within the $t-r$--plane in the tangent space. Therefore, we have
\begin{align}
	\label{Eq_}
  	(k^\mu) = (k^t,k^r, 0, 0) \, .
\end{align}

\begin{figure}
	\centering
	\includegraphics[width=0.49\textwidth]{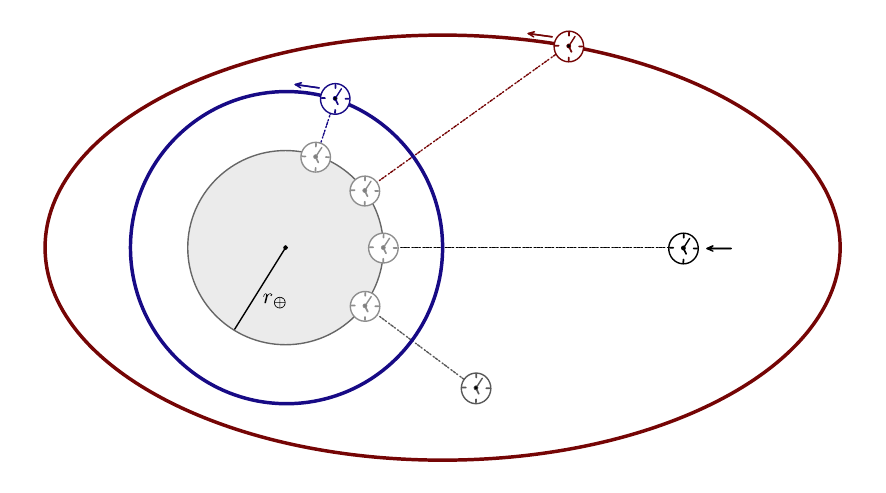}
	\caption{\label{Fig:comparison_scheme} Schematic description of the signal transmission: radial lightlike signals are sent within the equatorial plane. We assume a continuous distribution of receiver clocks along the Earth's equator with no mutual redshift. The clock comparison is done between the moving clocks on various orbits and the respective Earth--bound clock which is reached first by the radial lightlike signal.}
\end{figure}

For both observers, the four-velocities are normalized according to
\begin{align}
  		 g(u,u) \big|_{\gamma} = -1 \, , \quad g(\tilde{u}, \tilde{u}) \big|_{\tilde{\gamma}} = -1 \, ,
\end{align}
and their worldlines are, therefore, parametrized by their respective proper times. For the Earth--bound observers, we can immediately calculate the $t-$component of the four--velocity:
\begin{align}
	\label{Eq_}
  	g(\tilde{u}, \tilde{u}) \big|_{\tilde{\gamma}} = -f(r_\oplus) (\tilde{u}^t)^2 + r_\oplus^2 \Omega^2 = -1 \notag \\
  	 \Rightarrow \,\, \tilde{u}^t = \sqrt{\dfrac{1+r_\oplus^2\Omega^2}{f(r_\oplus)}} = \sqrt{\dfrac{1+v_\oplus^2}{f(r_\oplus)}} \, .
\end{align}
Above, we introduced $v_\oplus = r_\oplus \Omega^2$ as the rotational velocity of the Earth--bound clock. For $v_\oplus \equiv 0$, we recover the case of static observers fixed to the surface of a non--rotating Earth. 

We can now already give a general expression for the redshift between the Earth--bound observer and a second arbitrary observer that sends radial light signals for the frequency comparison. The result is
\begin{align}
	\label{Eq_redshift_general}
	1+z = \dfrac{\nu}{\tilde{\nu}} =\left( u^0(r,\varphi) + \dfrac{u^r(r,\varphi)}{f(r)} \right) \dfrac{\sqrt{f(r_\oplus)}}{\sqrt{1+v_\oplus}} \, ,
\end{align}
and is parametrized by the momentary position $(r,\varphi)$ at the emission of the signal.
Besides of being parametrized by proper time, no assumption is made so far on the emitter's worldline and four--velocity $u$. Hence, in Eq.\ \eqref{Eq_redshift_general}, $u^t$ and $u^r$ can be chosen also to correspond to an arbitrary (non--geodesic) motion. However, in the following we will mainly consider geodesic observers and specify these velocity components.

Since the emitter of the signal shall be a geodesic observer, the geodesic equation 
\begin{align}
	\ddot{x}^\mu + \Gamma^\mu{}_{\nu\sigma} \dot{x}^\nu \dot{x}^\sigma = 0
\end{align}
must be fulfilled along the worldline $\gamma$. The Lagrangian to describe the observer's motion is
\begin{align}
	\label{Eq_}
  	\mathcal{L} = \dfrac{1}{2} g(u,u) \, ,
\end{align}
and there are two constants of motion related to the two Killing vector field $\partial_t$ and $\partial_\varphi$
\begin{subequations}
	\begin{align}
		\label{Eq_}
  		E &:= f(r) u^t = -u_t \, , \\
  		L &:= r^2 u^\varphi = u_\varphi \, .
	\end{align}
\end{subequations}
The normalization of the four-velocity then contains the same information as the $r$-component of the Euler-Lagrange equations, and we obtain
\begin{align}
	\label{Eq_}
  	u^r = \pm \sqrt{E^2-\left(\dfrac{L^2}{r^2}+1\right) f(r)} \, .
\end{align}

For any geodesic light ray $\lambda$, however, there exist constants of motion as well
\begin{subequations}
	\begin{align}
		\label{Eq_}
  		E_{\lambda} &:= f(r) \, k^t = -k_t\, , \\
  		L_{\lambda} &:= r^2 k^\varphi = k_\varphi \, .
	\end{align}
\end{subequations}
For radial light rays in particular, which we consider for signal exchange between the observers, the angular momentum vanishes identically, $L_{\lambda}=0$. Since for any lightlike geodesic $g(k,k) = 0$ holds along the orbit, we further obtain
\begin{align}
	\label{Eq_}
  	k^r = \pm E_{\lambda} \, .
\end{align}
To summarize, we have in total for the respective four--velocities of the observers and the light ray's tangent vector
\begin{subequations}
\begin{align}
	\big(u^\mu \big) &= \left( E/f(r), \pm \sqrt{E^2-\left(\dfrac{L^2}{r^2}+1\right) f(r)}, L/r^2, 0 \right) \, , \\
	\big(\tilde{u}^\mu \big) &= \left( \sqrt{\dfrac{1+r_\oplus^2\Omega^2}{f(r_\oplus)}}, 0, \Omega, 0 \right) \, , \\
	\big(k^\mu \big) &= E_{\lambda} \, \left(f(r)^{-1}, \pm 1, 0, 0 \right) \, .
\end{align}
\end{subequations}

Note that we could also consider a co--rotating coordinate system to describe the rotating Earth and observers on its surface by the transformation $\varphi \to \bar{\varphi} = \varphi - \omega t$, where $\omega := \tilde{u}^\varphi/\tilde{u}^t$ is the angular frequency w.r.t.\ coordinate time. However, we choose to keep the non--rotating coordinates and we use these coordinates exclusively throughout the rest of the present paper.

\section{\label{Sec_redshift} Clock comparison and redshift}
In the following sections, we determine the redshift between an Earth--bound observer $\tilde{\gamma}$ and different (geodesic) observers $\gamma$ in orbits around the Earth. We consider static observers hovering in space, the case of free radial infall, circular geodesic motion, and arbitrary bound quasi--elliptical geodesic orbits. We determine the respective redshift for a radial signal transfer, see Fig.\ \ref{Fig:comparison_scheme} for a sketch of the situation. We assume a continuous distribution of receiver clocks along the Earth's equator with no mutual redshift since they move on an isochronometric surface \cite{Philipp2017} of the Schwarzschild spacetime and these Earth--bound clocks form an isometric congruence. The respective clock comparison is done between the moving clocks on various orbits and the respective Earth--bound clock which is reached first by the emitted radial light signal.

\subsection{Static observer}
This first scenario is an extension of the standard textbook example in which the redshift between two static observers is calculated, see, e.g., Ref.\ \cite{Wald2010}. 
Consider the Earth--bound observer at radius $r_\oplus$ and a second observer hovering at $r_0$. Assume that a radial light ray with tangent $k$ is emitted at the larger radius $r_0 > r_\oplus$ tangential to an $r$-line. The motion of the two observers and the light ray is then described by
\begin{subequations}
\begin{align}
	\big(\tilde{u}^\mu \big) &= \left( \sqrt{\dfrac{1+r_\oplus^2\Omega^2}{f(r_\oplus)}}, 0, \Omega, 0 \right) \, , \\
	\big(u^\mu \big) &= \left( 1/\sqrt{f(r_0)}, 0, 0, 0 \right) \, , \\
	\big(k^\mu \big) &= E_\lambda \left(f(r)^{-1}, -1 , 0, 0 \right) \, ,
\end{align}
\end{subequations}
respectively.
According to the general equations \eqref{Eq_Notation:RedshiftDef2} and \eqref{Eq_redshift_general}, the redshift between emitter and receiver is
\begin{align}
	\label{Eq_Result:zStatic}
	1+z_{\text{stat}} 
	= \dfrac{\nu}{\tilde{\nu}} 
	&= \sqrt{ \dfrac{f(r_\oplus)}{f(r_0)} } \dfrac{1}{\sqrt{1+r_\oplus^2\Omega^2}} \notag \\
	&= \sqrt{ \dfrac{f(r_\oplus)}{f(r_0)} } \dfrac{1}{\sqrt{1+v_\oplus^2}} \, .
\end{align}

\begin{figure}
	\includegraphics[width=0.45\textwidth]{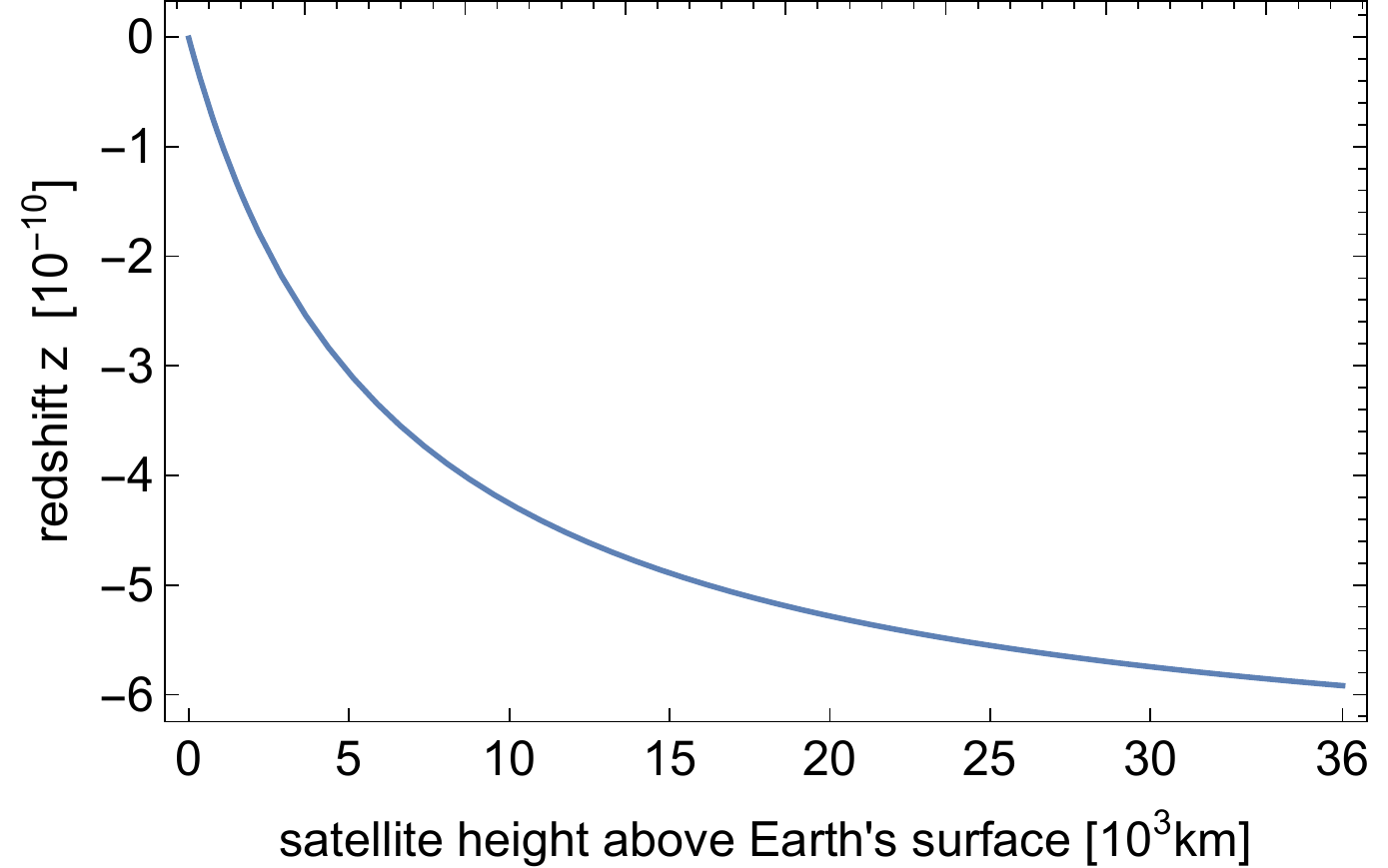}
	\caption{\label{Fig:redshift_stat} The redshift between an Earth--bound observer and an observer hovering at radius $r_0$. We show the redshift for different heights above the Earth's surface up to the geostationary height of about $36000\,$km. The total redshift contains the transverse Doppler contribution due to the Earth's rotation as well as the  gravitational effect.}
\end{figure}

Since $f(r)$ is a monotonically increasing function and $1/\sqrt{1+v_\oplus^2} < 1$, we have
\begin{align}
	 1+z_{\text{stat}} < 1 \, ,
\end{align}
and the redshift is negative. Hence, the frequency $\nu$ is smaller than $\tilde{\nu}$ and the clock at smaller radius 'runs slower'. In Fig.\ \ref{Fig:redshift_stat}, the redshift is shown as a function of the static satellite height $r_0$. The factor $(1+v_\oplus^2)^{-1/2}$ in Eq.\ \eqref{Eq_Result:zStatic} is related to the transverse Doppler contribution, already known in SR. It modulates the gravitational redshift and is caused by the circular motion of the Earth--bound receiver clock.

Calculating the first order post--Newtonian (weak--field) limit of Eq.\ \eqref{Eq_Result:zStatic}, we recover the well--known limit for the gravitational redshift and Doppler redshift in the post--Newtonian framework
\begin{align}
	\big( 1 + z_{\text{stat}} \big)_{\text{pN}} = \left( \dfrac{\nu}{\tilde{\nu}} \right)_{pN} = 1 + \dfrac{1}{c^2} \left( \Delta U - \dfrac{v_{\oplus,SI}^2}{2} \right) \, ,
\end{align}
see, e.g., Ref.\ \cite{Ashby2003} and references therein for a comparison. Here, we use $U(r) = -GM/r$, which is the Newtonian gravitational potential of a spherically symmetric source, and $\Delta U := U(r_\oplus) - U(r_0)$. For the limit procedure, we introduced SI--units and $v_{\oplus,SI} = c \, v_\oplus$ is the Earth--bound clock's velocity in [m/s].
We see that the gravitational redshift, to lowest order, is given by $\Delta U /c^2$. Redshift experiments are therefore sensitive to potential differences.
We can estimate the error made by using only the first order approximation above by considering the next order term in the week field expansion of $1+z_{\text{stat}}$, which is 
\begin{align}
	\dfrac{1}{c^4} \left( - \Delta U^2 \dfrac{r_0+3r_\oplus}{r_0-r_\oplus} - \dfrac{\Delta U v_{\oplus,SI}^2}{2} + \dfrac{3v_{\oplus,SI}^4}{8} \right) \, .
\end{align}
The last two terms are even several orders of magnitude smaller than the leading term which is proportional to $\Delta U^2/c^4$.
The entire second order contribution is evidently proportional to $c^{-4}$, and for a satellite height of about $5000\,$km above the Earth's surface, this contribution is approximately $-3 \times 10^{-19}$ and therefore 9 orders of magnitude smaller than the first order contribution, which is about $-3\times 10^{-10}$ for this situation. Hence, for redshift experiments sensitive to the $10^{-19}$--level, at least second order contributions must be taken into account. However, this exceeds the present accuracy by some orders of magnitude, see, e.g., \cite{Delva2015}.

\subsection{Radial infall}
\begin{figure}
	\includegraphics[width=0.45\textwidth]{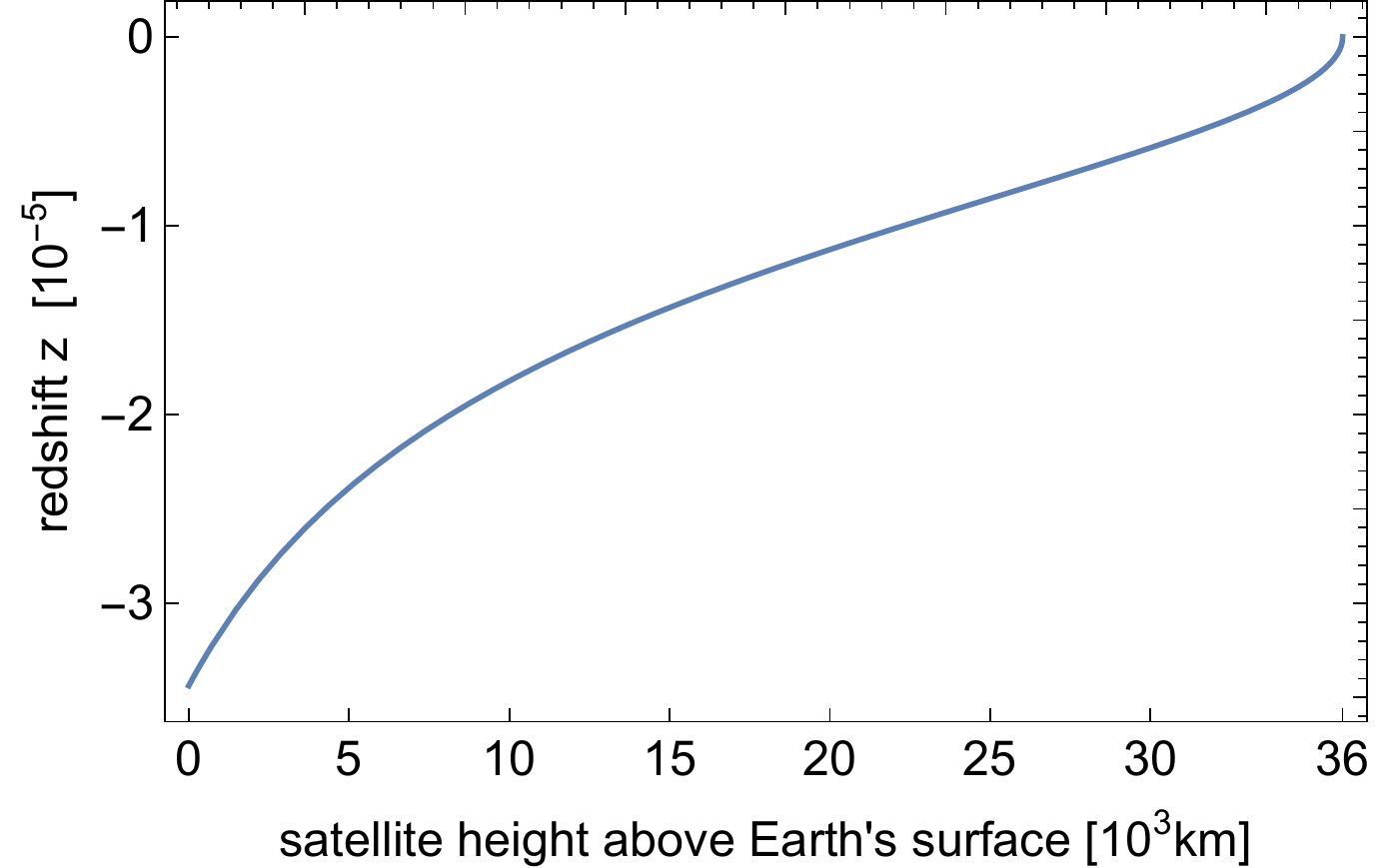}
	\caption{\label{Fig:redshift_infall}The redshift between an Earth--bound observer and an observer in radial infall at momentary height above the Earth's surface. The motion starts with zero initial velocity from a geostationary radius. The total redshift contains transverse and parallel Doppler contributions as well as the gravitational effect.}
\end{figure}

Assume now that the observer $\gamma$ is related to a satellite in radial free fall, starting at some initial position $r_0$ with initial velocity $\dot{r}_0$ parallel to an $r$--line. During this radial infall, light rays are emitted continuously and received by the Earth--bound observer. The motion of the infalling satellite is described by
\begin{align}
	\big(u^\mu\big) &= \left( E/f(r), - \sqrt{E^2-f(r)}, 0, 0 \right) \, ,
\end{align}
where $L=0$ holds since the angular momentum must vanish for radial motion. The radial position as a function of proper time is given by the solution of
\begin{align}
	\left( \dfrac{\mathrm{d}r}{\mathrm{d}\tau} \right)^2 = E^2 - f(r) \, .
\end{align}
The redshift between the infalling observer $\gamma$ and the observer $\tilde{\gamma}$ on the Earth, parametrized by the momentary radial position $r$ at the time of signal emission, is given by
\begin{align}
	1+z_{\text{rad}} &= \dfrac{\nu}{\tilde{\nu}} = \sqrt{ \dfrac{f(r_\oplus)}{f(r)}} \, \dfrac{E-\sqrt{E^2-f(r)}}{\sqrt{f(r)}\sqrt{1+v_\oplus^2}} \notag \\
	&= (1 + z_{stat} ) \, \dfrac{E-\sqrt{E^2-f(r)}}{\sqrt{f(r)}} \, ,
\end{align}
where the constant of motion $E$ is related to the initial conditions. Assuming that the motion of the infalling observer starts at a certain radius $r_0$ with initial radial velocity $\dot{r}_0$, we obtain
\begin{align}
	E = \sqrt{\dot{r}_0^2 + f(r_0)} \, .
\end{align}
In the Fig.\ \ref{Fig:redshift_infall}, we show an example of such a scenario, for which assume the radial infall to begin from a geostationary height with zero initial velocity $\dot{r}_0 = 0$.

When the satellite's motion starts from rest, a post--Newtonian expansion in inverse powers of $c$ leads to
\begin{align}
	\big( 1+z_{\text{rad}} \big)_{\text{pN}} = 1 - \dfrac{v_{\text{sat}}}{c} + \dfrac{1}{c^2} \left( \Delta U - \dfrac{v_{\oplus,SI}^2}{2} + \dfrac{v_{\text{sat}}^2}{2} \right) \, ,
\end{align}
where $v_{\text{sat}} := \sqrt{2GM/r - 2GM/r_0}$ is the (Newtonian) satellite's velocity at radius $r$ after being dropped from rest at radius $r_0$.
We recognize parallel Doppler effect terms, which are proportional to $v_{\text{sat}}/c$, and the gravitational contribution proportional to the potential difference $\Delta U/c^2$ between the Earth and the momentary satellite position. The transverse Doppler effect $\sim v_{\oplus,SI}^2/c^2$ is due to the Earth's rotation.

\subsection{Circular motion}
\begin{figure}
	\includegraphics[width=0.45\textwidth]{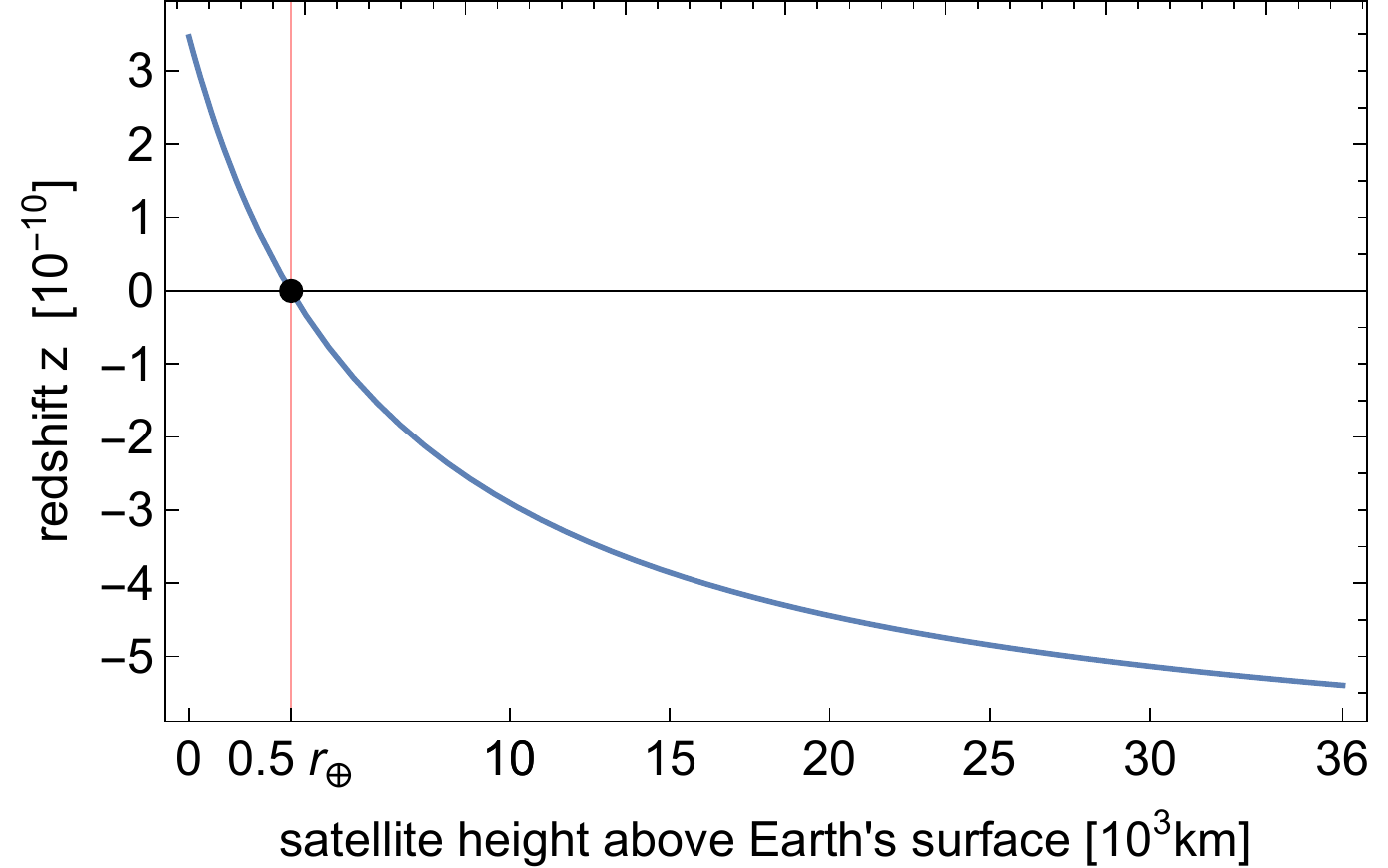}
	\caption{\label{Fig:redshift_circ} The redshift between an Earth--bound observer at $r_\oplus$ and an observer on a circular geodesic with radius $R$.}
\end{figure}

The next situation which we consider is the redshift between the Earth--bound observer $\tilde{\gamma}$ and another observer on a circular geodesic orbit with radius $r=R$.
The satellite's motion is described by
\begin{align}
	\big( u^\mu \big) = \left( \dfrac{E}{f(R)},0,\dfrac{L}{R^2},0 \right) \, ,
\end{align}
where for circular geodesics we have
\begin{subequations}
\begin{align}
	E^2 &= f(R)^2 \dfrac{R}{R-3m} \, , \\
	L^2 	&= \dfrac{m R^2}{R-3m} \, .
\end{align}
\end{subequations}
The redshift is
\begin{align}
	1+z_{\text{circ}} &= \dfrac{\nu}{\tilde{\nu}} = \sqrt{ \dfrac{f(r_\oplus)}{f(R)} } \, \dfrac{\sqrt{f(R)}}{\sqrt{1-3m/R}\sqrt{1+v_\oplus^2}} \notag \\
	&= (1 + z_{\text{stat}} ) \, \dfrac{\sqrt{f(R)}}{\sqrt{1-\dfrac{3m}{R}}}
\end{align}
In Fig.\ \ref{Fig:redshift_circ}, we show this redshift for circular geodesics with different radii.

Note that there exist a special radius given by ${r = 1.5\, r_\oplus \times \sqrt{1+v_\oplus^2}}$, for which the redshift vanishes identically due to cancellation of Doppler and gravitational redshift contributions.

A post--Newtonian expansion of the redshift result leads to
\begin{align}
	\big( 1+z_{\text{circ}} \big)_{\text{pN}} = 1 + \dfrac{1}{c^2} \left( \Delta U - \dfrac{v_{\oplus,SI}^2}{2} + \dfrac{v_\text{sat}^2}{2} \right) \, ,
\end{align}
where we can clearly see the gravitational redshift related to potential differences $\Delta U/c^2$ between the clock on Earth and the clock on the circular orbit, as well as the transverse Doppler effect which is proportional to $v^2/c^2$. Here, $v_\text{sat} = \sqrt{GM/r}$ is the Newtonian velocity of a satellite on a circular Kepler orbit.

\subsection{Arbitrary geodesic motion}
Now, we consider an observer $\gamma$ that moves on an arbitrary bound orbit and evaluate the redshift w.r.t.\ the Earth--bound observer $\tilde{\gamma}$. 
We investigate in particular i) a nearly circular orbit around the Earth with a small eccentricity, ii) the Schwarzschild analog of the orbit of the ``mislaunched'' Galileo satellite 5 with a moderate eccentricity of $e=0.15$, iii) a highly elliptic orbit around the Earth with eccentricity $e=0.6$, and iv) an elliptic orbit around a black hole that clearly shows perigee precession.

The redshift between the Earth--bound clock and the moving geodesic observer is
\begin{align}
	\label{Eq_redshift_elliptic}
	1+z_{\text{elliptic}} &= \dfrac{\nu}{\tilde{\nu}} \notag \\
	&= \sqrt{ \dfrac{f(r_\oplus)}{f(r)} } \, \dfrac{E \pm \sqrt{E^2 - (L^2/r^2 +1)f(r)}}{\sqrt{f(r)}\sqrt{1+v_\oplus^2}} \notag \\
	&= (1 + z_{\text{stat}} ) \, \dfrac{E \pm \sqrt{E^2 - (L^2/r^2 +1)f(r)}}{\sqrt{f(r)}} \, ,
\end{align}
where $E$ and $L$ are the constants of motion along $\gamma$. They are related to the initial conditions by
\begin{subequations}
	\begin{align}
	\label{Eq_SchwarzschildInitialConditions}
  	L 	&= r_0^2 \, \dot{\varphi}_0 \\
 	E^2 &= \dot{r}_0^2 + \left( 1- \dfrac{2m}{r_0} \right) \left( 1 + \dfrac{L^2}{r_0^2} \right) \, .
	\end{align}
\end{subequations}
We can also express $L$ and $E$ by a suitable defined semi--major axis and eccentricity of the orbit. The orbit has turning point $r_p$ (perigee) and $r_a$ (apogee). At these turning points, $\mathrm{d}r/\mathrm{d}\tau = 0$, and we can define eccentricity $e$ and semi-major axis $a$ by
\begin{align}
	r_p =: (1-e) a \, , \quad r_a =: (1+e) a \, .
\end{align}
Now, we can relate $E$ and $L$ to $e$ and $a$ by
\begin{subequations}
\label{Eq_Schwarzschild_orbitalparam}
\begin{align}
	L^2 &= \dfrac{ f(r_p) - f(r_a) }{ \frac{f(r_a)}{r_a^2} - \frac{f(r_p)}{r_p^2} } \, , \\
	E^2 &= \left( \dfrac{ L^2 }{ r_a^2 } + 1 \right) f(r_a) \, .
\end{align}
\end{subequations}
Inserting this result into Eq.\ \eqref{Eq_redshift_elliptic} yields the redshift as a function of the momentary radial position, the eccentricity, and the semi--major axis. The radial position, for any chosen combination of $e$ and $a$, can only vary between $r_p$ and $r_a$.

We argue that, however, Eq.\ \eqref{Eq_redshift_elliptic} might still be misleading, because the redshift $z_{\text{elliptic}}$ is given as a function of the momentary radial position $r$ at the signal's emission. For an arbitrary elliptic orbit, all radial values between perigee and apogee are realized, but for a better understanding we can insert the solution of the equation of motion to obtain the redshift as a function of proper time or the azimuthal angle. This highlights in particular the periodic character of the redshift signal. An analytic solution of the geodesic equation is
\begin{align}
	\label{Eq_WeierstrassSolution}
	r(\varphi) = \dfrac{m}{2\powerset(\varphi - \varphi_{in}) + 1/6} \, ,
\end{align}
where $\varphi_{\text{in}}$ is related to the initial conditions according to
\begin{align}
	\varphi_{\text{in}} = \varphi_0 + \int_{y_0}^{\infty} \dfrac{\mathrm{d}z}{\sqrt{4z^3-g_2z-g_3}} \, , \, y_0 = \dfrac{1}{2}\left(\dfrac{m}{r_0} - \dfrac{1}{6} \right) \, .
\end{align}
The Weierstrass invariants $g_2$ and $g_3$ are determined by the constants of motion as follows:
\begin{subequations}
	\begin{align}
		g_2 &= \dfrac{1}{12} - \dfrac{m^2}{L^2} \, , \\
		g_2 &= \dfrac{1}{216} - \dfrac{1}{12}\dfrac{m^2}{L^2} - \dfrac{1}{4}\dfrac{m^2}{L^2}\left(E^2-1\right) \, .
	\end{align}
\end{subequations}
For details on the analytic solution and possible applications, we refer the reader to the seminal paper by Hagihara \cite{Hagihara1930} and the work in Refs.\ \cite{Hackmann2008,Hackmann2008b,Hackmann2009}. In Ref.\ \cite{Hackmann2012} observables for bound orbits in more general axially symmetric spacetimes are considered, and in the present work, we add the exact description of the relativistic redshift for the Schwarzschild spacetime as a possible observable, see Eq.\ \eqref{Eq_redshift_elliptic}.

Inserting \eqref{Eq_WeierstrassSolution} into \eqref{Eq_redshift_elliptic} yields the redshift as a function of the azimuthal angle as an alternative.

The post--Newtonian limit of Eq.\ \eqref{Eq_redshift_elliptic} leads to
\begin{align}
	\label{Eq_redshift_elliptic_pN}
	\left( 1+z_{\text{elliptic}} \right)_{\text{pN}} = 1 + \dfrac{v_{\text{sat}\,\parallel}}{c} +\dfrac{1}{c^2} \left( \Delta U - \dfrac{v_{\oplus,SI}^2}{2} + \dfrac{v_\text{sat}^2}{2} \right) \, ,
\end{align}
where the satellites (Newtonian) orbital velocity is 
\begin{align}
	v_\text{sat} = \sqrt{G M \left( \dfrac{2}{r} - \dfrac{1}{a} \right)} \, ,
\end{align}
and the parallel velocity component turns out to be 
\begin{align}
	\label{Eq_satellite_velocity_parallel1}
	v_{\text{sat}\,\parallel} = \sqrt{\dfrac{GM}{a} \dfrac{(r-r_a)(r_p-r)}{r^2}} \, .
\end{align}
Using relations for Newtonian Kepler orbits, it can also be expressed in the well known form
\begin{align}
	\label{Eq_satellite_velocity_parallel2}
	v_{\text{sat}\,\parallel} = \sqrt{\dfrac{G M}{p}} e \sin \theta \, ,
\end{align}
where $p := a(1-e^2)$ is the semilatus rectum and $\theta$ is the true anomaly. In the expansion \eqref{Eq_redshift_elliptic_pN}, apart from the genuine gravitational redshift $\sim \Delta U/c^2$, we recognize parallel Doppler effects proportional to $v_{\text{sat}\,\parallel}/c$ as well as transverse Doppler terms proportional to $v/c^2$ due to the circular motion of the Earth--bound clock as well as the transverse velocity of the satellite. It can easily be deduced from the relations \eqref{Eq_satellite_velocity_parallel1} and \eqref{Eq_satellite_velocity_parallel2} for the parallel velocity component that the parallel Doppler effect vanishes at the turning points $r=r_a$ and $r=r_p$ as we expect it to bee according to the special relativistic formulae.

\begin{figure*}
	\includegraphics[width=0.85\textwidth]{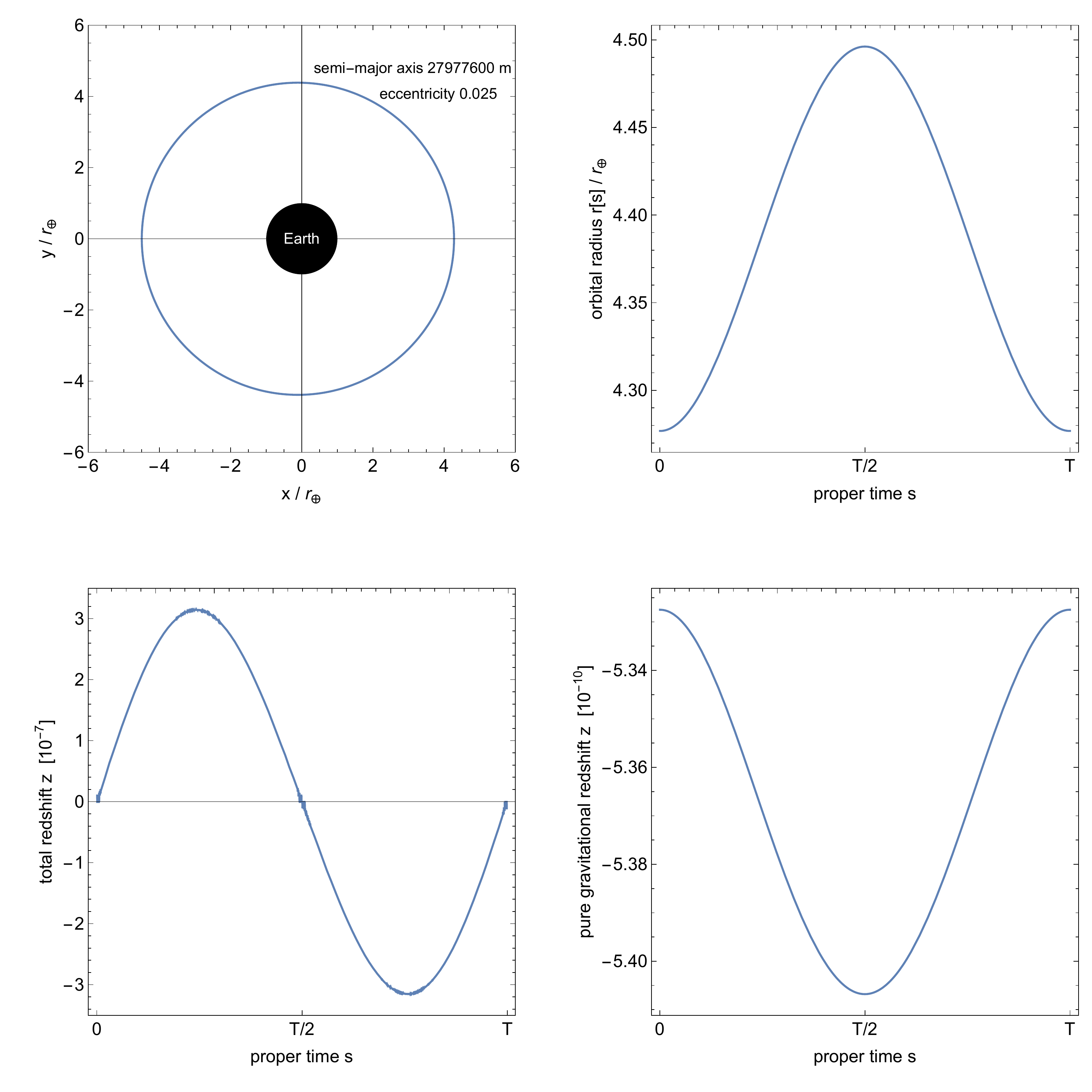}
	\caption{\label{Fig_ACirc} Redshift between an almost circular orbit with eccentricity $e=0.025$ and semi major axis $a=27977600\,$m and an observer on the surface of the rotating Earth. We show the orbit as well as the radial coordinate $r$ over one full orbital period (upper row). In the bottom row, we show the total and gravitational redshift, respectively. The total redshift contains besides the gravitational contribution, transverse and parallel Doppler terms. The peak--to--peak difference in the gravitational redshift is about $1 \times 10^{-12}$.}
\end{figure*}

\begin{figure*}
	\includegraphics[width=0.85\textwidth]{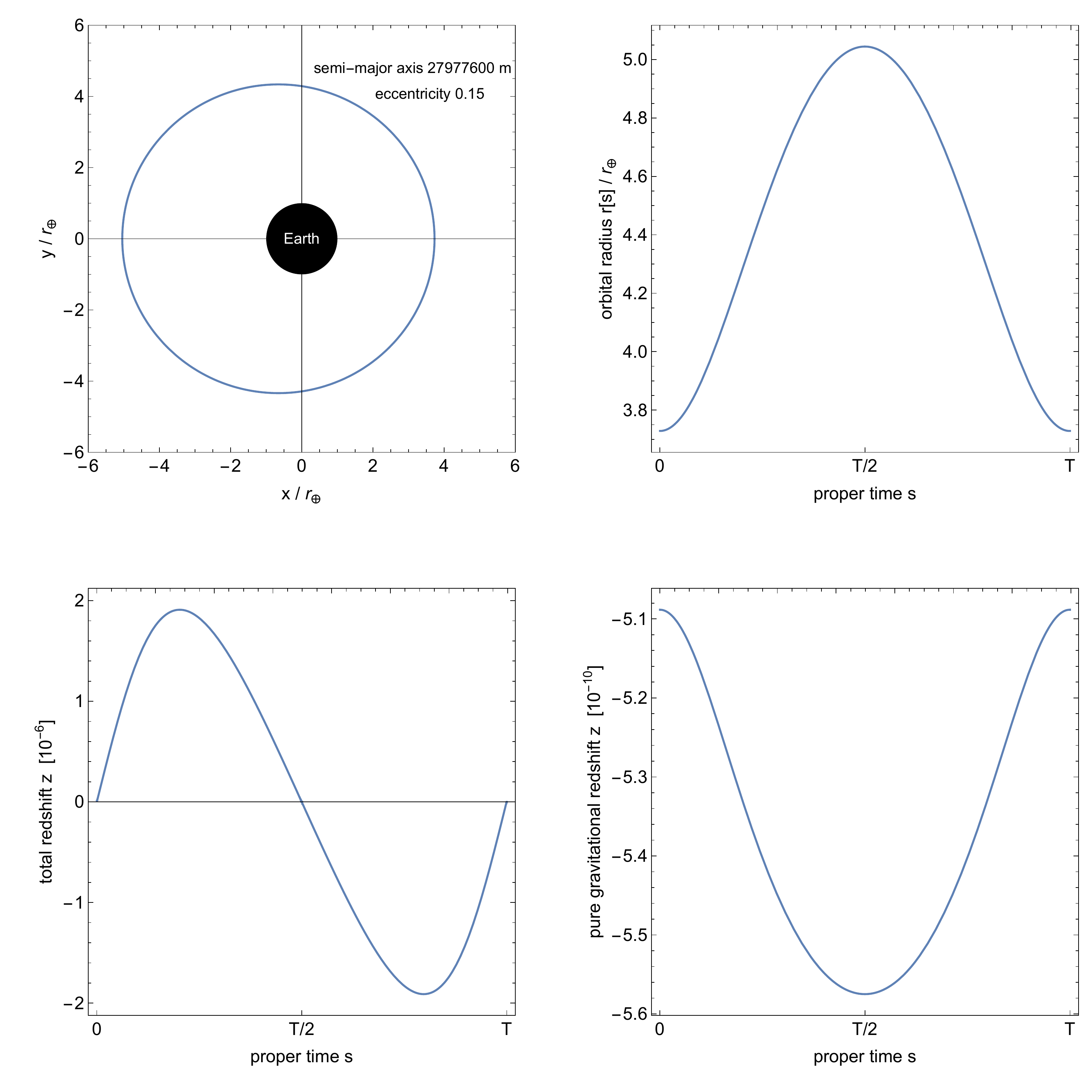}
	\caption{\label{Fig_Galileo} Redshift between the Schwarzschild version of the Galileo 5 orbit with an orbital eccentricity $e=0.15$ and semi major axis $a=27977600\,$m and an observer on the surface of the rotating Earth. We show the orbit as well as the radial coordinate $r$ over one full orbital period (upper row). In the bottom row, we show the total and gravitational redshift, respectively. The total redshift contains besides the gravitational contribution, transverse and parallel Doppler terms. The peak--to--peak difference in the gravitational redshift is about $5 \times 10^{-11}$.}
\end{figure*}

\begin{figure*}
	\includegraphics[width=0.85\textwidth]{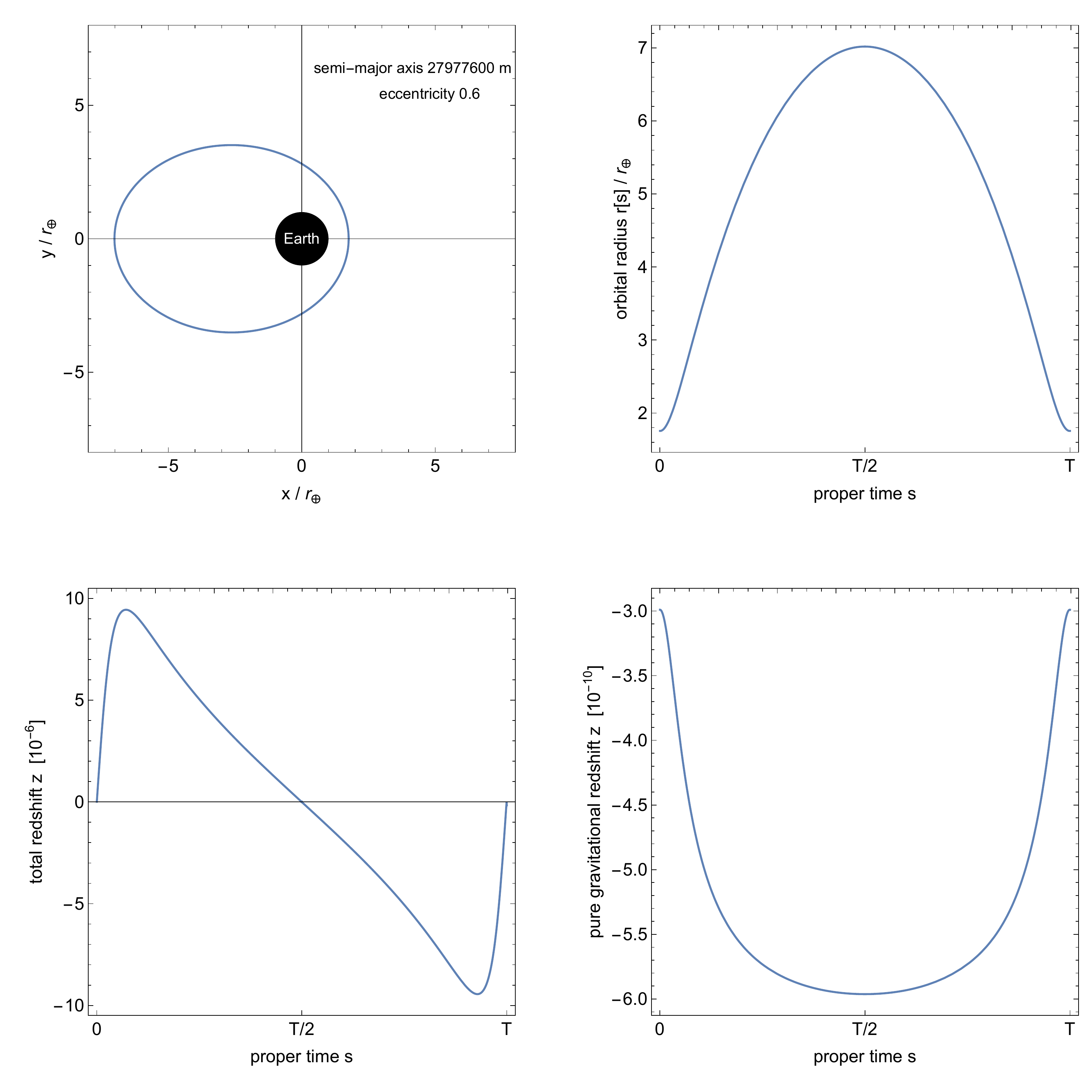}
	\caption{\label{Fig_Elliptic} Redshift between a highly elliptical orbit with eccentricity $e=0.6$ and semi major axis $a=27977600\,$m and an observer on the surface of the rotating Earth. We show the orbit as well as the radial coordinate $r$ over one full orbital period (upper row). In the bottom row, we show the total and gravitational redshift, respectively. The total redshift contains besides the gravitational contribution, transverse and parallel Doppler terms. The peak--to--peak difference in the gravitational redshift is about $3 \times 10^{-10}$.}
\end{figure*}

\begin{figure*}
	\includegraphics[width=0.85\textwidth]{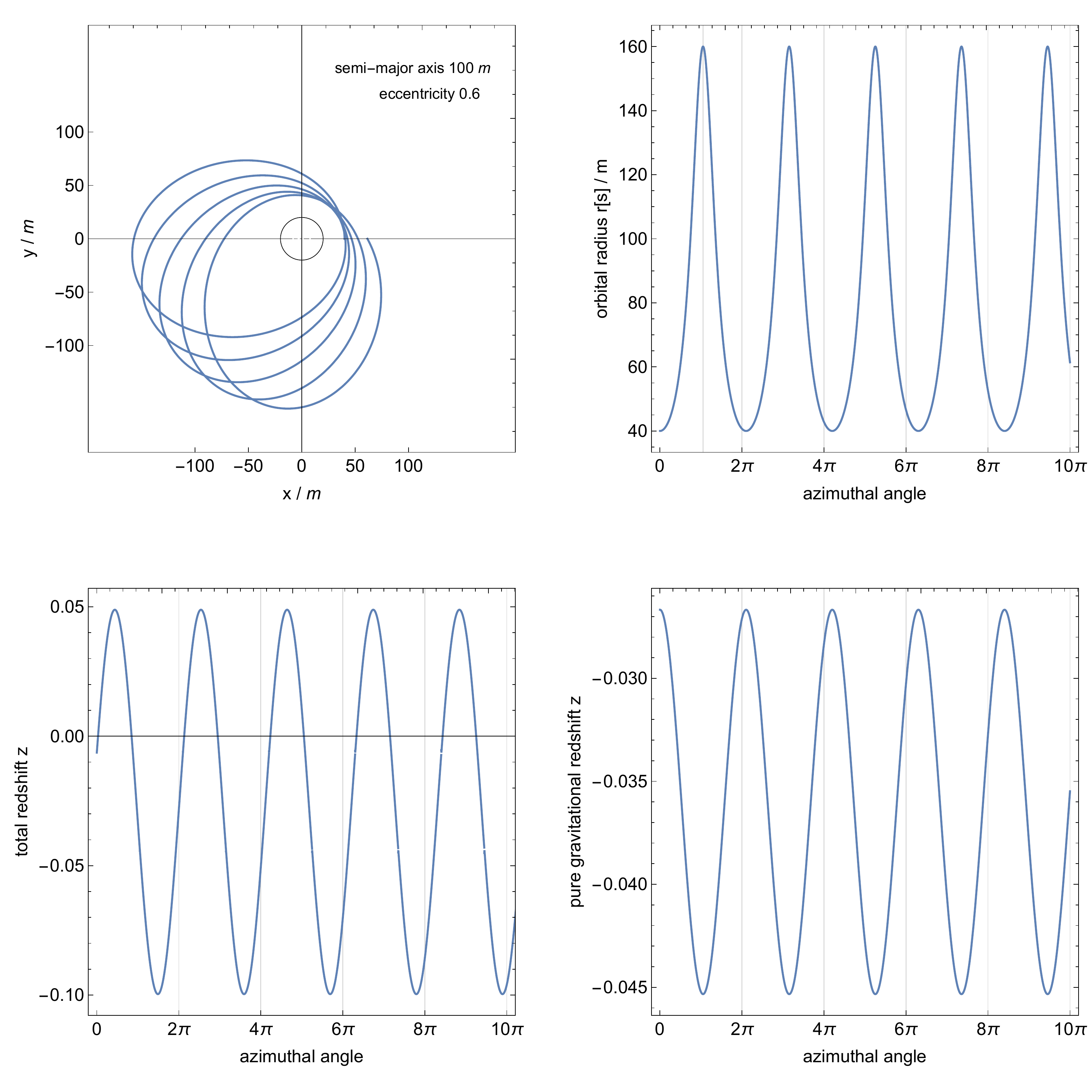}
	\caption{\label{Fig_BH} Redshift between a highly elliptical orbit with eccentricity $e=0.6$ and semi major $100\, m$ around a Schwarzschild black hole and static observers at $r=20\, m$. We show the orbit as well as the radial coordinate $r$ over one full orbital period (upper row). The black circle in the left plot indicates the position of the static observers that receive the signals. In the bottom row, we show the total and  gravitational redshift, respectively. The total redshift contains besides the gravitational contribution, transverse and parallel Doppler terms.}
\end{figure*}

As a possible application of our results, we consider now four different elliptical bound orbits. These are i) an almost circular orbit around the Earth with a moderate eccentricity $e=0.025$, ii) a Schwarzschild version of the Galileo 5 orbit with an eccentricity of $e=0.15$, iii) a highly elliptical orbit around the Earth with $e=0.6$, and iv) a highly elliptical orbit with $e=0.6$ and semi--major axis of $100\,m$ around a Schwarzschild black hole. The results for the total as well as the gravitational redshift are shown in Figs.\ \ref{Fig_ACirc} -- \ref{Fig_BH}.

The gravitational redshift takes its largest value at the apogee since it is the largest distance and therefore the largest potential difference $\Delta U$. In Eq.\ \eqref{Eq_redshift_elliptic_pN}, we have shown that the gravitational redshift is, to lowest order, sensitive to these potential differences. Figs.\ \ref{Fig_ACirc} -- \ref{Fig_Elliptic} show how the profile of the gravitational redshift widens with increasing eccentricity. However, the gravitational redshift is 3--4 orders of magnitude smaller than the total redshift due to the large Doppler contributions. Hence, besides an accurate clock comparison, a precise knowledge of the satellite's state vector is needed to recover the gravitational redshift from experimental data.

Investigating the peak--to--peak differences for the gravitational redshift in the three scenarios shown in Fig.\ \ref{Fig_ACirc} -- \ref{Fig_Elliptic}, we observe the following relations:\\

\begin{tabular}{l|cc}
  orbital eccentricity & peak--to--peak grav.\ redshift\\
  \hline
  \hline \\
  0.025 & $\approx 1 \times 10^{-12}$ \\
  0.15 & $\approx 5 \times 10^{-11}$ \\
  0.6 & $\approx 3 \times 10^{-10}$ \\
\end{tabular}\\\\

Fig.\ \ref{Fig_BH} shows the relation between the redshift and the periapsis precession. We have plotted several orbits and drawn vertical lines at multiples of $2\pi$ to guide the eye. The perigee, i.e.\ the minimal radial position shifts after each orbit. At the same time, also the apoapsis precesses by the same angle and we see this shift phase--locked with the maxima and minima of the gravitational redshift.


\section{\label{Sec_multipoles} Higher order multipole effects}
So far, we have used the Schwarzschild spacetime, which possesses only a monopole moment, to calculate the redshift between two observers. Within the theory of GR, we can extend this framework to consider also the influence of higher order multipole moments on the gravitational redshift. In the following, we will consider static observers in a general Weyl spacetime and, in particular, focus on the influence of the quadrupole moment on the gravitational redshift.

The Weyl class of spacetimes contains all static, axisymmetric, and asymptotically flat solutions of Einstein's vacuum field equation \cite{Weyl1917}. The metric of a general Weyl spacetime is
\begin{multline}
	\label{Eq_WeylMetricSpheroidal}
	g_{\mu \nu} dx^{\mu} dx^{\nu}  = -e^{2\psi} dt^2 + m^2 e^{-2\psi} (x^2-1)(1-y^2)d\varphi^2
	\\
	+ m^2 e^{-2\psi} e^{2\gamma} (x^2-y^2) 
	\left( \dfrac{dx^2}{x^2-1} + \dfrac{dy^2}{1-y^2} \right) \, ,
\end{multline}
where we use a time coordinate $t$, an azimuthal coordinate $\varphi$ adapted to the symmetry, and spheroidal coordinates $(x,y)$, see \cite{Quevedo1989}. The metric function $\psi$ is given by the expansion
\begin{align}
	\label{Eq_WeylMetricExpansionSpheroidal}
	\psi(x,y) = \sum_{l=0}^\infty (-1)^{l+1} q_l \, Q_l(x) \, P_l(y) \, ,	
\end{align} 
where $P_l(y)$ are Legendre polynomials and the $Q_l(x)$ are Legendre functions of the second kind as given in, e.g., Ref.\ \cite{Bateman1955}. The parameters $q_l$ are related to the Newtonian multipole moments of an axisymmetric gravitational potential in the Newtonian limit, as shown in Refs.\ \cite{Quevedo1989} and \cite{Philipp2017}. We can relate the spheroidal coordinates $(x,y)$ to quasi--spherical (Schwarzschild--like) coordinates $(r,\vartheta)$ by the transformation
\begin{align}
\label{Eq_CoordinateTrafo}
	x:=r/m-1 \, , \quad y := \cos \vartheta \, .
\end{align}
If only $q_0 \neq 0$ and all other $q_l$ for $l>1$ vanish identically, the Schwarzschild spacetime is recovered.

Now, we consider two static observers with four--velocities $u$ and $\tilde{u}$, respectively, in a Weyl spacetime. For such static observers we have
\begin{align}
	\big( u^\mu \big) = \big( u^t, 0 , 0 ,0 \big) \, , \quad \big( \tilde{u}^\mu \big) = \big( \tilde{u}^t, 0 , 0 ,0 \big) \, .
\end{align}
The normalization of the four--velocity immediately yields
\begin{align}
	u^t = \mathrm{e}^{-\psi(x,y)} \, , \quad \tilde{u}^t = \mathrm{e}^{-\psi(\tilde{x},\tilde{y})} \, .
\end{align}
According to the general equation \eqref{Eq_Notation:RedshiftDef1}, the redshift between these two static observers is
\begin{align}
	\label{Eq_redshiftWeyl1}
	1+z_{\text{Weyl}} = \dfrac{\nu}{\tilde{\nu}} = \dfrac{u^t}{\tilde{u}^t} = \dfrac{\mathrm{e}^{\psi(\tilde{x},\tilde{y})}}{\mathrm{e}^{\psi(x,y)}} \, ,
\end{align}
where $(\tilde{x},\tilde{y})$ and $(x,y)$ are the observers' positions, respectively.
Note, however, that static observers in a Weyl spacetime are on surfaces of constant redshift potential. Hence, the redshift \eqref{Eq_redshiftWeyl1} between any two of these isochronometric surfaces can easily be calculated using the framework of the redshift potential, see \cite{Philipp2017}. Inserting the expansion \eqref{Eq_WeylMetricExpansionSpheroidal} for the Weyl metric function, we finally obtain for the redshift
\begin{align}
	1+z_{\text{Weyl}} = \dfrac{\exp \left( \sum_{l=0}^N (-1)^{l+1} q_l Q_l(\tilde{x}) P_l(\tilde{y}) \right) }{\exp \left( \sum_{l=0}^N (-1)^{l+1} q_l Q_l(x) P_l(y) \right)} \, .
\end{align}
This expression is valid for all Weyl spacetimes with multipole moments up to order $N$ in the Newtonian limit, and it is exact, i.e.\ no approximations are involved. For the choice $q_0=1$ and $q_l = 0$ for all $l>0$, we recover the Schwarzschild result of the previous section \ref{Sec_redshift}.

In Ref.\ \cite{Philipp2017} it is shown how the Newtonian limit of the metric function $\psi$ is calculated. Based on these calculations, we obtain here the post--Newtonian expression of the redshift above, which is
\begin{align}
	\label{Eq_redshiftWeylPN}
	\big(1+z_{\text{Weyl}}\big)_{\text{pN}} = 1 + \dfrac{\Delta U}{c^2} \, ,
\end{align}
where 
\begin{align}
	U = -G \, \sum_{l=0}^N N_l \dfrac{P_l(\cos \vartheta)}{r^{l+1}} \, ,
\end{align}
which is the Newtonian gravitational potential of an axisymmetric mass distribution with Newtonian multipole moments $N_l$. The $N_l$ are then related to the $q_l$ via
\begin{align}
	N_l = (-1)^{l} q_l \dfrac{l!}{(2l+1)!!} (G/c^2)^{l} \, M^{l+1} \, .
\end{align}

Eq.\ \eqref{Eq_redshiftWeylPN} demonstrates that also for a more general multipolar mass distribution, the post--Newtonian expression for the gravitational redshift between static observers is, to lowest order, determined by potential differences.


\subsection{The quadrupole contribution}
Besides the monopole, the next order multipole contribution to the gravitational redshift comes from the quadrupole. To treat such a configuration in full GR, we choose $q_0=1, \, q_2 \neq 0$ and all other $q_l=0$. This choice yields the Erez-Rosen spacetime, where the metric potential is given by
\begin{multline}
	2 \psi = \log \left( \dfrac{x-1}{x+1} \right) + q_2 (3y^2-1)\left( \dfrac{(3x^2-1)}{4} \right. \\
	\times \left. \log \left( \dfrac{x-1}{x+1} \right) + \dfrac{3}{2} x \right) \, .
\end{multline}
This spacetime is a quadrupolar generalization of the Schwarzschild solution. We introduce the Schwarzschild--like spherical coordinates by \eqref{Eq_CoordinateTrafo}. Then, the exact gravitational redshift for two static observers in such a quadrupolar spacetime becomes
\begin{align}
	\label{Eq_redshiftErez-Rosen}
	1+z_{\text{quadr}} = \sqrt{\dfrac{f(r_\oplus)}{f(r)}} \dfrac{h(r_\oplus,\vartheta_\oplus)}{h(r,\vartheta)} \, .
\end{align}

One of the two observers shall be located on the surface of the Earth, i.e.\ at $(r_\oplus,\vartheta_\oplus)$, and the second static observer is located in space at $(r,\vartheta)$.
The function $f(r)$ is the Schwarzschild metric function and $h(r,\vartheta)$ is given by
\begin{multline}
	h(r,\vartheta) = \exp \left \lbrace q_2 (3\cos^2\vartheta-1) \times \left[ \left( \dfrac{3}{4} \left(\dfrac{r}{m}-1 \right)^2- \dfrac{1}{4} \right) \right. \right. \\
	\left. \left. \times \log \left( 1-\dfrac{2m}{r} \right) + \dfrac{3}{2} \left(\dfrac{r}{m}-1\right) \right] \right \rbrace \, .
\end{multline}
The first term in \eqref{Eq_redshiftErez-Rosen} resembles the Schwarzschild result for a monopolar gravitational field, and the term $h(r_\oplus,\vartheta_\oplus)/h(r,\vartheta)$  describes the modification of the gravitational redshift due to the general relativistic quadrupole.

Calculating the post--Newtonian limit of the result \eqref{Eq_redshiftErez-Rosen} we obtain
\begin{align}
	\big( 1+z_{\text{quadr}}\big)_{\text{pN}} &= 1 + \dfrac{1}{c^2} \left( -\dfrac{GM}{r_\oplus} - G N_2 \dfrac{3\cos^2 \vartheta_\oplus - 1 }{2r^3} \right. \notag \\
	&\left. + \dfrac{GM}{r} G N_2 \dfrac{3\cos^2 \vartheta - 1 }{2r^3} \right) \notag \\
	&= 1 + \dfrac{\Delta U_{\text{quadr}}}{c^2} \, ,
\end{align}
where the value of $q_2$ is chosen to be related to the Earth's Newtonian quadrupole
\begin{align}
	q_2 = \dfrac{15}{2} \dfrac{N_2}{M m^2} \, .
\end{align}
Instead of $N_2$, sometimes the dimensionless parameter
\begin{align}
	J_2 = \dfrac{N_2}{a_\oplus^2 M}
\end{align}
is used. Here, $a_\oplus$ is the Earth's mean radius. For the Earth, the value of $J_2$ is approximately ${J_2 \approx 1.0826\times 10^{-3}}$. Hence, the quadrupolar contribution to the Newtonian gravitational potential is three orders of magnitude smaller than the monopolar contribution due to the total mass $M$. Our result in the post--Newtonian limit is in agreement with the well--known equations for the gravitational redshift beyond spherically symmetric gravitational fields, see, e.g., Ref.\ \cite{Ashby2003}.


\section{Conclusion}

Within the framework of General Relativity, we have derived an exact analytic expression for the redshift between an Earth--bound receiver clock and arbitrary observers in space that emit radial light signals for the frequency comparison in Schwarzschild spacetime. 
Furthermore, we have specified the results to the case of a static observer hovering at a constant spatial position, geodesic observers in radial free fall, in circular orbit, as well as in general bound elliptical motion. 

The post--Newtonian expansions of our results reveal the contributions to the total redshift due to the transverse and parallel Doppler effects, as well as the genuine gravitational redshift, which is related to potential differences. 
The transverse Doppler contributions are due to the circular motion of the receiver clock on the Earth and the perpendicular (w.r.t.\ the orbital trajectory) velocity of the satellite that emits the electromagnetic signal for the frequency comparison. 
The parallel Doppler effect appears whenever the satellite has a non--vanishing radial velocity and it vanishes at the turning points of bound orbits as shown by the respective equations Sec.\ \ref{Sec_redshift}.

For the redshift between an observer on the surface of the rotating Earth and satellites in bound quasi--elliptical geodesic motion, we have considered three cases with different eccentricities. We have shown how the profile of the gravitational redshift widens with increasing eccentricity and that Doppler effects on the redshift are usually a few orders of magnitude larger. Hence, the satellite's state vector must be known accurately to deduce the gravitational redshift from experimental data in high--precision frequency or clock comparison experiments.

To analyze the influence of relativistic higher order multipoles, we have derived an exact expression for the gravitational redshift for static observers in a general Weyl spacetime.
In particular, we have shown how the relativistic quadrupole modifies the Schwarzschild result.
The Newtonian limit of our result for the gravitational redshift was related to potential differences of axisymmetric Newtonian gravitational potentials in lowest order.

In our future work, we will extend our framework to cover also non--radial signal transmission and gravitomagnetic contributions to the redshift and clock comparison by considering rotating spacetimes, such as the Kerr spacetime, and moving observers in these geometries. 
Furthermore, we will analytically and without approximations treat the influence of higher order relativistic multipole moments on the Doppler and gravitational redshift by considering for instance Weyl spacetimes and Quevedo--Mashhoon spacetimes and the mutual redshift also for moving observers. These axisymmetric static spacetimes possess well--defined Newtonian limits, see Sec.\ \ref{Sec_multipoles}, and this allows also to recover the influence of multipole moments in the post--Newtonian framework.
It has to be analyzed how our exact results for the relativistic redshift can contribute to future satellite experiments and data analysis. However, for high--precision measurements, environmental effects on the satellite orbits need to be taken into account, and the satellite's motion is no longer described by unperturbed Kepler--like or Schwarzschild geodesics. Even though a first order post--Newtonian treatment of the relativistic redshift might be sufficient in some situation, contemporary and future measurements should be accompanied with the best available theoretical framework in full GR to ensure an undoubtedly correct interpretation.



\begin{acknowledgments}
This work was supported by the Deutsche Forschungsgemeinschaft (DFG) through the Collaborative Research Center (SFB) 1128 ``geo-Q'' and the Research Training Group 1620 ``Models of Gravity''. We also acknowledge support by the German Space Agency DLR and the RELAGAL study with funds provided by the Federal Ministry of Economics and Technology (BMWi) under grant number DLR 50WM1547.
\end{acknowledgments}

\clearpage



%

\end{document}